\documentclass[12pt]{article}
\usepackage[margin=0.9in]{geometry}
\usepackage{float}
\usepackage{amsmath}
\usepackage{amssymb}
\usepackage{graphicx}
\usepackage{cite}
\usepackage{color}
\usepackage{dcolumn}
\usepackage{bm}

\begin{document}

\begin{flushleft}
{\large
\textbf{Understanding cancer complexome using
networks, spectral graph theory and multilayer framework}
}
\\
Aparna Rai$^{1}$,
Priodyuti Pradhan$^{2}$,
Jyothi Nagraj$^{3}$,
K. Lohitesh$^{3}$,
Rajdeep Chowdhury$^{3}$,
Sarika Jalan$^{1, 2, \ast}$
 \\
\bf{1} Centre for Biosciences and Biomedical Engineering, Indian Institute of Technology Indore, Simrol, Indore, Madhya Pradesh 453552, India
\\
\bf{2} Complex Systems Lab, Discipline of Physics, Indian Institute of Technology Indore, Simrol, Indore, Madhya Pradesh 453552, India
\\
\bf{3} Department of Biological Sciences, Birla Institute of Technology and Science, Vidya Vihar, Pilani, Rajasthan 333031, India
\\
$\ast$ Corresponding Author (Email: sarikajalan9@gmail.com)
\end{flushleft}

\begin{abstract}

Cancer complexome comprises a heterogeneous and multifactorial milieu that varies in cytology, physiology, signaling mechanisms and response to therapy. The combined framework of network theory and spectral graph theory along with the multilayer analysis provides a comprehensive approach to analyze the proteomic data of seven different cancers, namely, breast, oral, ovarian, cervical, lung, colon and prostate.
Our analysis demonstrates that the protein-protein interaction networks of the normal and the cancerous tissues associated with the seven cancers have overall similar structural and spectral properties. However, few of these properties implicate unsystematic changes from the normal to the disease networks depicting difference in the interactions and highlighting changes in the complexity of different cancers. 
Importantly, analysis of common proteins of all the cancer networks reveals few proteins namely the sensors, which not only occupy significant position in all the layers but also have direct involvement in causing cancer. The prediction and analysis of miRNAs targeting these sensor proteins hint towards the possible role of these proteins in tumorigenesis. This novel approach helps in understanding cancer at the fundamental level and provides a clue to develop promising and nascent concept of single drug therapy for multiple diseases as well as personalized medicine.

\end{abstract}


The post-genomic era aims to understand human health and diseases 
by investigating the role of proteomics and genomics, that involves macromolecules such as the proteins and nucleic acids (e.g. DNA, RNA, miRNA etc) \cite{Venter}. 
Cancer being a multifactorial disease can be studied through these macro-molecules.
To understand this complexome, there has been rapid advancements in both experimental and 
theoretical techniques in cancer diagnostic and screening \cite{6}.
These investigations indicate that all the cancers share a common pathogenic 
mechanism \cite{2002}.  Much like Darwinian evolution, cancer cells acquire
continuous heritable genetic variation by arrays of random mutation and 
go through the process of natural selection resulting in phenotypic diversity \cite{Marusyk, Garraway} 
like, 
differential gene expressions, alterations in cell regulation and control mechanisms, 
alteration in macromolecular interaction pathways, etc.
These two fundamental processes in cancer cells provide them the capacity to have proliferative advantage 
and higher rate of survival than their neighboring cells \cite{Marusyk} resulting in 
heterogeneous tumor formations \cite{Stratton}. 
This heterogeneity is found in both intra- and inter-tumor cell populations \cite{Fisher}. 
In addition, there are non-genetic factors that result in phenotypic diversity, 
e.g. epigenetic modifications, clinical diagnostic and therapeutic responses \cite{Burrell, Epigenetic_Portela}.
All these factors result to aberrations in various biological processes of the cancer cells 
and make cancer a complexome with no direct correspondence between the cancer and the
normal tissues.
These studies have remarkably improved our understanding of various factors 
associated with the cancer. However, even after billion dollars of investments \cite{3, 33} and extensive research, the major challenge lies in understanding
the angiogenic and metastatic complexity \cite{4}, modeling the disease at a global scale, drug target
identification and co-evolving tumor cell \cite{5}. These challenges forms the backbone of 
cancer systems biology.  The research involving genetics at the molecular level 
have identified a number of susceptible genes responsible for the genesis of different 
types of cancers \cite{7}. However, out of about
10$\%$ of the total cancer genes only 1$\%$ are known for their biological functions, 
indicating that the etiology of cancer is still not clear \cite{8}.
This demands for a holistic approach to understand cancer from a fundamental point of view.
One such promising approach is to consider the cancer system as networks.
Studying cancer under network theory framework has already helped in 
understanding various constitutive properties of the system 
\cite{barabasi_2004, Draghici, CancerNetBalkwill}.
Various network studies pertaining to, epigenetic modifications, gene regulations, 
gene expressions, protein-protein interactions (PPI) have provided insights into the 
molecular mechanisms of the disease. Additionally, these network studies
have helped in finding functionally important proteins 
as well as some of the missing pathways in cancer \cite{wang, cancer01, cancer02, cancer03}
providing a global understanding to biological processes and protein interactions \cite{Creixell, barabasi, cancer1, cancer2, cancer3}.

In this work, we consider seven most prevalent in human cancers namely, the breast, oral, ovarian, cervical, lung, colon and prostate. We analyze the 
protein-protein interactions among the
normal and disease cells using the combined framework of network theory, spectral graph theory and the multilayer analysis
to understand cancer development, progression and treatment response. 
The spectral graph theory has shown its remarkable success in uncovering the 
behavior of various complex systems \cite{cancer3, rmt_sj, rmt0, rmt_sj1, rmt1, spectra_camellia, rmt_sj2, rmt3} including biological 
systems corresponding to gene co-expression and PPI networks \cite{gene_sj, rmt2, rmt4, spectra_pramod}.
Further, implementing the multilayer analysis, we scale these seven cancers on the basis of the presence of common proteins 
into three different categories elaborately discussed in the methods section. 
The network approach of analyzing seven cancers not only demonstrates 
deviation in the complexity of all the normal and disease datasets from their 
corresponding random networks 
but also depicts changes in the complexity level between the normal and the disease states,
contributing to understand cancer at the fundamental level.
The multilayer framework highlights the proteins which are common 
in all the cancers and have  
structural importance in individual networks. Importantly, these common proteins
 also exhibit functional importance for occurrence of cancer
revealed through pathway ontology and miRNA analysis. The framework paves a new way to the promising and nascent concept of
single drug therapy \cite{10} for multiple diseases as well as personalized medicine \cite{11}
in a time efficient and cost effective manner.

\section*{Results and discussions}
\subsection*{Properties of Complexome}

\begin{figure}[ht]
\begin{center}
\includegraphics[width=12cm,height=8cm]{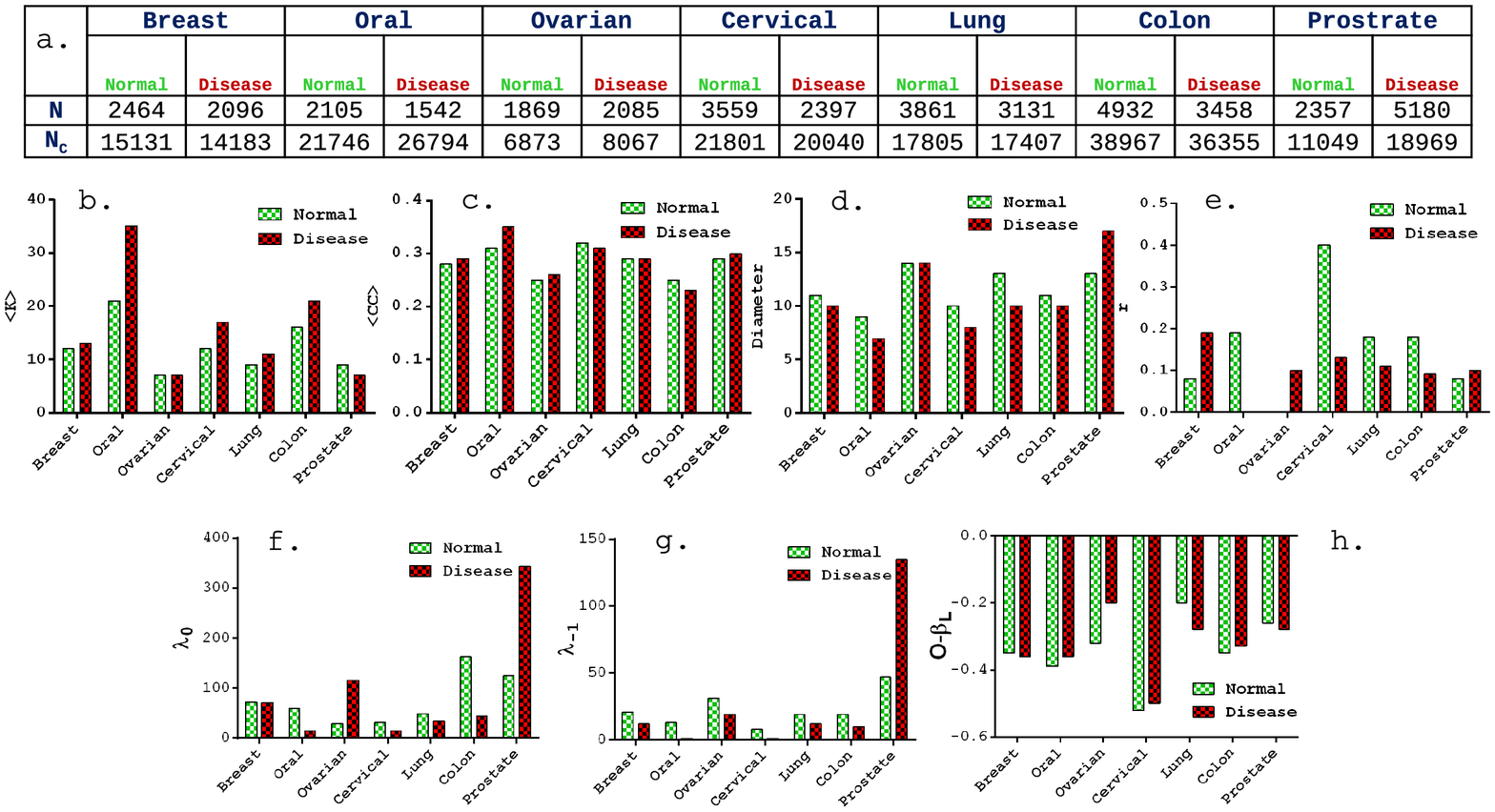}
\end{center}
\caption{{\bf Different structural and spectral properties.} The table (a) summaries number of nodes ($N$) and number of connections ($N_C$) of all PPI networks. The graph represents (b) the average degree $\langle k \rangle$, (c) average clustering coefficient $\langle C \rangle$, (d) diameter $D$, (e) degree-degree coefficient $r$, degenerative eigenvalues: (f) $\lambda_{-1}$ and (g) $\lambda_{0}$, and 
(g) betweenness-overlap correlation ($O-\beta_{L}$) for the real and their corresponding random networks for normal and disease datasets.
All the cancers show similar statistics for $\langle CC \rangle$, diameter, $\lambda_{-1}$, $\lambda_0$, except in prostate cancer. 
There is no significant comparison in the values of $\langle CC \rangle$ and $r$ between normal and corresponding disease networks.} 
\label{Fig_properties}
\end{figure}

\paragraph*{Structural Properties}
We determine various structural properties of all the seven cancers for the normal and the 
disease states (Table. S1 and S2). 
Additionally, we perform the comparative analysis of various properties of these networks 
with those of their corresponding random networks. We construct the corresponding 
Erd\"os R\'enyi (ER) random networks with the same $N$ and average degree as of the PPI 
networks. The ER network only preserve the N and $\langle k \rangle$, but have interactions among pairs randomly chosen. 
Another random network model which we consider here is the configuration model. The configuration model additionally preserves the degree sequence of a network for the normal and the disease states.
As indicated from Fig. \ref{Fig_properties}a, all the disease datasets consist of 
different number of proteins as compared to their individual normal datasets. It is well 
implicated in various cancer proteome related studies that advantageous genetics and 
natural selection \cite{Stratton} lead to mutations in the proteins and impaired functions,
causing different count of proteins participating in cancer networks \cite{Beerenwinkel}. 

To get an overview of 
the structural organization of the protein interactions, we first examine the global structural properties of the
normal and the disease networks.
First such property is the number of connections.
As indicated in Fig. \ref{Fig_properties}b, the overall average degree is higher in the disease networks except for the ovarian and prostate cancer, 
indicating more tendency of proteins to interact with each other in these cancers. 
Other PPI studies have also reported that most of the cancer proteins exhibit 
a higher binding tendency to interact with other proteins due to favorable mutations at binding sites \cite{cancer02}.
Hence, the observation of higher number of interactions in the disease state is not surprising. 
In the seven cancers considered here, the disease network of oral cancer exhibits relatively a higher number of connections than that of all the other
networks suggesting that for this particular cancer, proteins have more affinity to interact among themselves. The observation is already reported that they have large molecular organizations facilitating maximum interactions with other proteins \cite{oral_degree}.
Further, the degree distribution for all these PPI networks follow power law behavior which is 
also observed for many other biological systems \cite{barabasi_2004}. 
More intriguingly, the degree distribution
of the all the networks 
follow two distinct fitting scales, i.e. two power law for all the
networks (Fig. S1). Many real world networks have been reported
to devoid of a perfect power law for the entire range of the degree 
\cite{double_powerlaw, pramod_epl2015}. The existence of two power law implicates that the network is robust not only by the inclusion of new proteins and interactions by the hub proteins but also by the contribution of new or altered interactions of existing proteins \cite{double_powerlaw}.

Second such global structural property is diameter ($D$) of the network which captures the spread of information in a system \cite{diameter}. Diameter of a network is the longest of the shortest paths between all the pair of nodes of the network \cite{diameter}. 
Different networks having same values of $N$ and $N_C$ (number of connections) can have different diameter \cite{Watts}. Similarly, networks having different values of $N$ and $N_C$ can yield the same diameter as diameter is decided by the manner in which connections are distributed in a network. 
A small diameter is known for the faster signaling in the system \cite{Watts}. Based on the diameter analysis, we have two results; first the diameter of PPI networks is larger than that of their corresponding ER and configuration random models (Table. S2). 
Second, the disease PPI networks display relatively smaller diameter as compared to the corresponding normal ones. The smaller diameter in the disease state may be due to both the smaller size as well as large average degree of the disease networks (Table. S2).
Since, diameter has been emphasized to shed light on information propagation in a network \cite{Watts}, a small diameter of the disease network indicates a faster propagation of signals in terms of less number of the intermediate paths involved. 
It is already known that disorders in cancer associated proteins promote the adaptability of 
faster communication in many major cancer related cellular signaling processes by up 
regulation or down regulation of pathways leading to uncontrolled cell proliferation \cite{Csermely, diameter_cancer}. 
Short path for PPI networks can also be considered with respect to time as few molecules 
such as miRNAs facilitating the disease proteins to
regulate their expressions by mediating inter-cellular cell signaling in cancer cells which further lead to 
faster information transduction within the cell \cite{diameter_time}.
Note that, here networks are constructed by taking PPI
from the same database (STRING) for both the normal and disease states and hence these networks do not capture dynamical behavior of the PPI's.
A possible way of capturing the dynamical nature of PPIs is to consider gene-co-expression networks \cite{gene_1, gene_2} etc., but gene-co-expression networks also have their own limitations and drawbacks such as small sample size \cite{gene_3}. 

Third global property indicating the interaction complexity of networks
is the degree-degree correlation coefficient ($r$) which indicates 
tendency of nodes to connect with (dis)similar degree nodes \cite{Alok_r}.
Almost all the networks possess positive $r$ values except for the 
oral disease and the ovarian normal dataset. Since the $r$ values of these two
datasets is very close to zero, we can consider them to be neutral.
Overall, both the disease and the normal PPI networks display positive or near to zero 
values of $r$, whereas the 
corresponding ER networks have $r$ value zero.
This is not surprising as the networks with the same average degree and size may still differ significantly in various network features based on their nature of interactions. In the ER random network, the nodes are randomly connected and $r$ value of the network is determined by degrees of the interacting nodes.
Further, in order to see whether changes in the degree-degree correlations of different 
networks are arising due to the change in the degree sequence, we compare all the PPI 
networks with their corresponding configuration networks. The $r$ value of the fourteen 
configuration networks turns out to be overall dissortative in nature. Therefore, some normal networks are more assortative than the disease
networks while, vice-versa behavior is seen for some cancers (Fig. \ref{Fig_properties}e).
Since, ER random and configuration model  depict similar
behavior of all the fourteen networks, the changes in the $r$ values of
the networks constructed from the datasets have the following implications.
There are changes in the interaction 
patterns of the disease from the normal networks as the change in $r$ is not arising
due to
 changes in the degree sequence or data
size. 
This further implicates that there exists varying complexity in the 
underlying system. 
Complexity of molecular associations in different normal tissues can be understood 
as there are different proteins and pathways that define cell proteome and various 
sub-tissue types which are functioning for particular tissue fate.
Similarly, Cancer genetics and histopathology comprises of 
-intra and -inter tumor heterogeneity and its metastasis may harbor overall divergent 
biological behavior. In other words, different cancers may have altogether different 
cancer genetics and histopathology in their tumors \cite{Fisher}.
Altogether, the unsystematic changes in the values of $r$ from the normal to the disease networks 
indicate the inherent complexity of the two states.

Analysis of above global properties of the network indicate deviations of the real networks from their random counterparts
as well as reflects overall change in the complexity of interaction patterns
of the normal and the disease networks.
We further investigate properties providing understanding of
the local interactions in the network. One such property is the clustering coefficient of nodes 
in a network which suggests how the neighbors of a node are connected \cite{Watts, config}. 
The average clustering coefficient $(\langle CC \rangle)$ depicts an unsystematic change in its values in the disease and their normal networks, 
i.e. some disease datasets have more $\langle CC \rangle$, while for some datasets normal networks have more
$\langle CC \rangle$ (Fig. \ref{Fig_properties}c). Different values of $\langle CC \rangle$ further indicates changes in the interaction patterns in the normal and disease states.
But what is noteworthy is that these networks have much higher $\langle CC \rangle$ than those of their corresponding ER and configuration networks, 
indicating the abundance of cliques of order three \cite{ravraz_modules} .
Cliques indicate the preserved interactions in the networks and are believed to be conserved during the process of evolution \cite{Alon}. 
Further, these structures are also considered to be building blocks of a network, making the 
underlying system more robust \cite{Lotem} and stable \cite{Dwivedi} and cancer as a system is reported to be robust 
against both the targeted chemotherapy and the hazardous environment \cite{Kitano_Cancer, plos}.
Another property revealing structural complexity of a network is the correlation 
between degree and clustering coefficient ($k-CC$) for the nodes in a network.
The $k-CC$ correlations of all the disease and normal networks manifest overall negative value (Fig. S2), as 
also exhibited by many other biological systems reflecting the presence of hierarchical structure in the system \cite{barabasi_2004}. 
The presence of hierarchical structures is an indication of highly clustered neighborhoods consisting of sparsely connected nodes
communicating through hubs and functional modules in the network \cite{ravraz_modules}.

All the structural properties discussed above reveal overall similar behavior 
for almost all the normal and the disease states as well as indicate  existence of complexity in interaction pattern of both the networks . 
Further, the variation in the values of $r$ and $\langle CC \rangle$ from the normal to the disease,
i.e. in some of the datasets normal having more values of $r$ and $\langle CC \rangle$ than those of disease and vice versa, which reflect an unsystematic change in the interactions from the normal to the disease networks.
To have 
an in-depth analysis of proteins or factors causing changes in the interaction patterns in 
the PPI networks considered here, we explore the spectral properties  of these networks.

\paragraph*{Spectral properties}

The eigenvalue distributions of all the normal and disease networks depict triangular shape 
(Fig. S3) as observed for many other biological networks \cite{Aguiar}. 
Furthermore, the spectra exhibit a very high degeneracy at the zero eigenvalue for all the networks as compared to that of their corresponding ER random networks. The corresponding configuration models exhibit a high degeneracy at the zero eigenvalue which indicates that not only a particular degree sequence but also the nature by which these protein-protein interactions contribute on occurrence of the high degeneracy at the zero eigenvalues in the real networks.
Since number of zero eigenvalues in the adjacency matrix is directly related with the 
complete and partial node duplicates in the underlying network \cite{Alok}, 
a very high value of $\lambda_0$ degeneracy
indicates occurrence of node duplication in these PPI networks.
The duplicate nodes are the ones which shares the same neighbors
in a given network. Here we consider the nodes which are complete duplicates, that is,
these nodes have exactly the same neighbors. There are partial duplicate nodes also in the network
which do not have exactly the same neighbors but possess few uncommon neighbors too. Finding partial duplicate nodes in a network
is computationally very exhaustive and hence here we only concentrate on complete duplicate nodes.
These duplicate nodes are known to be important during the evolution \cite{Kitano, Golub} 
and hence occurrence of node duplication in the normal networks is not surprising as it 
indicates the evolutionary processes over the years. Interestingly, duplication is also known to be one of the important factor in promoting cancer and contribute in evolving the normal cell to the disease state \cite{dup_bio1, dup_bio2}.
But, what is interesting is that despite a very high $\lambda_{0}$ degeneracy in 
the disease networks, indicating a very high number of exact and partial duplicates, 
most of the duplicate nodes in the disease 
PPI networks are different from those of the corresponding normal PPI networks (Fig. S4).
This observation suggests that the genetic mutations leading to abrupt transformation 
of a normal cell into a cancerous cell may have caused incurrence of 
new proteins to perform similar functions and thus resulting in the duplicate nodes in the disease state.
Moreover, there is a different number of the duplicate nodes in different cancers which can be understood 
in terms of independent adaptation of each cancer genome arising due to independent heterogeneity 
and natural selection \cite{hallmark}. 
Further, both the disease and the normal networks display the degeneracy at minus one 
eigenvalues ($\lambda_{-1}$)  
whereas $\lambda_{-1}$ is absent in both the corresponding ER and the configuration networks.
It indicates that the PPI networks may contain complete sub-graphs.
Complete sub-graphs are known to be the building blocks of a network 
further making the network robust \cite{Lotem} and stable \cite{Dwivedi}.
It is known that architecture of PPI network is made up of sub-networks of metabolic cycles 
and pathways which play important role in constituting PPI networks \cite{SubGraphs, SubGraphs1}.

One striking observation is that both the disease and the normal datasets of prostate cancer 
result in different network properties than those of the other cancer datasets. For instance, 
some of the properties analyzed here, like the average degree $\langle k \rangle$, diameter and minus one degenerate eigenvalues of the prostate cancer are exception to that of the other networks.
The disease network of prostate has much higher value of zero eigenvalues than the normal, whereas for other pairs the vice versa behavior is observed with an exception of
ovarian cancer. For other properties such as the average clustering coefficient and degree-degree correlation coefficient, there is unsystematic changes from normal to disease.
Different behavior of prostate cancer may be arising 
due to lack of availability of complete knowledgebase of proteomic interactions for prostate cancer.
Further, it has been reported that prostate cancer is very lately diagnosed \cite{prostate} 
and thus, the altered network properties of prostate cancer may also
suggest the significance of independent cancer development processes in this cancer.

\subsection*{Multilayer analysis} 
\begin{figure}[ht]
\begin{center}
\includegraphics[width=9cm,height=4cm]{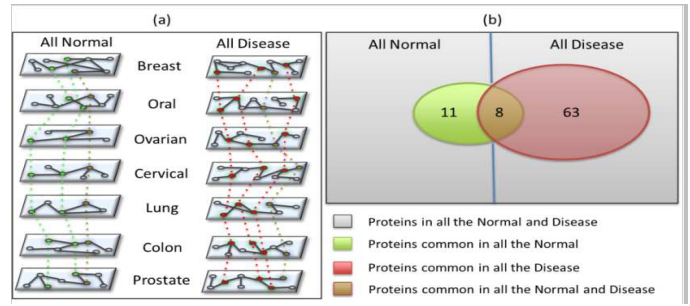}
\end{center}
\caption{{\bf Multilayer analysis.} (a) Schematic diagram showing the construction of multilayer network where each normal and disease network of the seven cancers are represented as layers leading to normal and disease multilayer networks respectively. The dotted lines represent the common proteins considered from each of dataset, the red, green and blue circles represent common proteins in all (i) the disease, (ii) the normal datasets and both the normal and disease datasets (union of (i) and (ii)), respectively. After extraction of these common proteins, their interaction partners are taken from individual datasets. 
(b) The Venn diagram of common proteins depicting the number of proteins common in all the normal and disease dataset, respectively.} 
\label{Fig2_multilayer}
\end{figure}

All the structural and spectral properties reveal similar behavior for almost all the seven normal and the disease networks 
such as high value of $\langle CC \rangle$, smaller diameter as compared to their corresponding random networks, 
non-negative $r$ values, negative $k-CC$ correlations and the triangular distribution of the eigenvalues with very few exceptions for each which have been discussed in the above section.
However, the disease state differ from their normal counterpart in terms of $r$ values
and $\langle CC \rangle$, suggesting difference in the interaction patterns between the disease and the normal networks.
To get insight into the proteins responsible for the changes in the interaction patterns from the
normal to the disease, which may also be crucial in making a
normal tissue to the cancerous one, we enlist the common proteins 
in all the normal as well as disease datasets (Fig. \ref{Fig2_multilayer}) as explained in the multilayer framework in the methods section. 
There are 63 proteins which are common in all the disease networks. If a protein appears in all the disease dataset, it is enlisted in the common proteins list.  
We investigate the pathways involved by considering the interactions 
of these common proteins as well as their structural importance in the networks.
There are 19 proteins which appear common in all the seven normal datasets and 71 in all the seven disease ones.
Out of these, 8 proteins appear common in all the normal and all the disease datasets (Fig. \ref{Fig2_multilayer}b).
The common proteins occur in different cancers as
the tumor cells share similar cellular environment
biologically i.e. uncontrolled cell division, cell proliferation, metastasis etc.
There are some reports on the occurrence of common protein markers in different 
cancers \cite{Hallmarks}
which suggest that there might be similar features associated with these common proteins present in different cancers.
We investigate the network properties of these common proteins for all the disease states
to find their importance in the network architecture
and also study their biological functions.

First, we extract the interacting partners of all these common proteins from the individual 
disease networks (Table. S5 and S6). We find that though the 
proteins are common in these PPI networks, some of the
interacting partners of these common proteins are different in individual networks 
suggesting addition or deletion of proteins due to mutations caused in each cancer.
Thereafter, we enlist the number of interactions among the 63 common proteins (referred as IN connections) and
the interactions outside the 63 proteins (referred as OUT connections) (Table. S6).  
These proteins have much higher number of OUT connections than the average degree of the
corresponding network 
(about two fold of average degree), suggesting their significant contribution in the overall network connectivity. 
We further analyze interaction properties of these subtractive PPI networks.
We determine the $\langle CC \rangle$ of these 63 nodes and compare it with
that of the whole network for all the seven disease datasets.
The $\langle CC \rangle$ of 63 nodes in each disease is much higher 
(nearly twice) as compared to the corresponding whole networks. 
A high $CC$ of 63 proteins indicates accountability of these proteins 
for higher $\langle CC \rangle$ of the whole network as well as existence of modular structures in the network.   
To have a broader understanding into the organization of these 63 proteins,
we perform molecular and pathway ontology of these common proteins.
To do this, we retrieve protein sequences of 63 proteins from UniProtKB and direct them to Reactome pathway browser \cite{Reactome}.
We find that, among the 63 common proteins, many proteins collectively play significant role in few of the important pathways such as 
Vascular endothelial growth factor (VEGF) signaling, PIP3 activation, 
VEGFA-VEGFR2 Pathway, Signaling by PDGF, Signaling by NGF etc (Table \ref{63_prtns}). 
All of these pathways are reported to have constitutive role in different cancers (detailed description in Supplementary Material). 

Furthermore, we investigate other structural properties such as 
$k-\beta_{c}$ (degree-betweenness centrality) correlation, weak ties analysis etc of these 63 proteins. The analysis reveals 
proteins contributing to the occurrence of the disease.

\subsection*{$k-\beta_{c}$ correlation and weak ties analysis}
\begin{figure}[ht]
\begin{center}
\includegraphics[width=12cm,height=6cm]{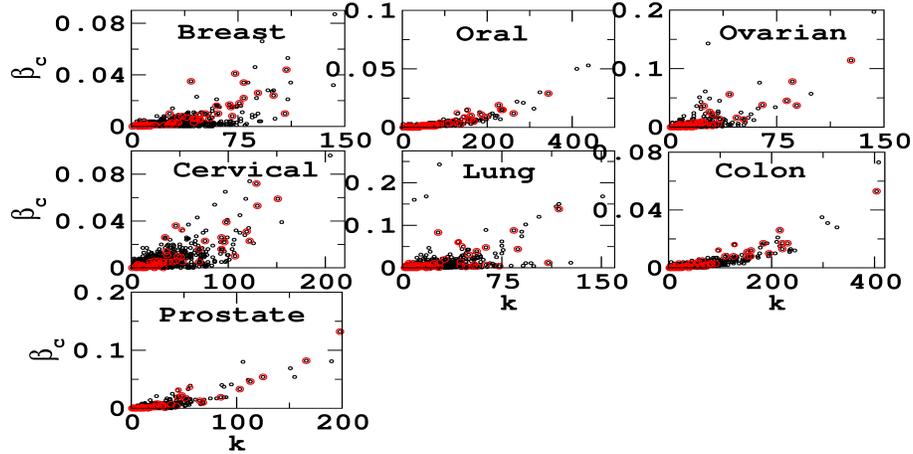}
\end{center}
\caption{{\bf $k-\beta_c$ correlation.} The $k-\beta_c$ correlation for all the disease networks reveal positive correlation (black circles). The red circles depict the $k-\beta_c$ correlation for 63 disease common networks.} 
\label{Fig_k_bc}
\end{figure}

\paragraph*{$k-\beta_{c}$ correlation}
We analyze the correlation between the degree and betweenness centrality for all the disease networks 
and highlight the 63 proteins as shown in Fig. \ref{Fig_k_bc}. 
All the disease networks exhibit overall positive $k-\beta_{c}$ correlation 
as also observed for many other complex systems \cite{barabasi_2004}.
Although few networks depicts data being clustered, i.e., data-points are localized in several regions, the overall correlation behavior of these networks are positive. Further,
there are few nodes which despite of having
less number of interactions (low $k$), participate in a large number of pathways calculated through
the betweenness centrality (high $\beta_{c}$). 
In biological context, these proteins may be important as 
betweenness measures the ways in which signals can pass through the interaction network.
If a protein having a high $\beta_{c}$ has low value of $k$, it depicts the participation of that 
protein in many pathways and connecting different functional modules. 
We find that there are many proteins having high $\beta_{c}$ and low $k$ in the individual disease datasets. 
However, here we consider only 63 disease common proteins which fall in this regime. 
Among 63 disease common proteins, six proteins possess remarkably high $\beta_{c}$ and low $k$ values in four (breast, ovarian, cervical and lung) of the seven disease networks. These proteins are namely
MUC1, STAT1, SOD2, MAPK1, HSPA4 and HSPA5. In the other three disease networks 
(oral, colon and prostate) these proteins possess high $\beta_{c}$, but comparative to other four networks they do not have very low $k$.  It is crucial to note here 
that these six proteins are not the hub proteins. The hub proteins have already been 
reported in carrying out many necessary and housekeeping functions in the cell \cite{hub},
and the list of significantly very high degree proteins or hub proteins can be found in Table. S3.
Let us focus on the six proteins revealed through $\beta_c$ and $k$ analysis
having structural importance in the network architecture. We look for the biological functions of these proteins
and find that these proteins are well implicated in cancers.
All these six proteins,   
are involved in the anti-apoptotic pathways \cite{apoptosis}. Particularly, five 
proteins, namely MUC1, STAT1, SOD2, MAPK1 and HSPA5 
are additionally responsible to aggravate metastasis \cite{metastasis}  
and play a key role in tumor progression by acting as angiogenic regulators \cite{angiogenesis}.
Further, three of these proteins STAT1, MAPK1 and HSPA5 also induce cell proliferation \cite{proliferation}.
Thus, the proteins having high $\beta_{c}$ and low degree exhibit 
significant involvement in cancer related activities.
The detailed functional properties are summarized in the supplementary material.

After investigating the nodes which are important in various pathways, 
we direct our attention towards finding important links or edges in the network connecting the 63 proteins, through Granovetter's `weak ties hypothesis'.

\paragraph*{Weak ties analysis}

As defined in the methods section, weak ties
are the links with their end nodes having very less number of common neighbors \cite{Granovetter}. 
This analysis, motivated from the social network research, highlights  importance of an edge in a network through its edge-betweenness centrality \cite{camellia_weak_ties}. 
An inverse relation between the strength of a tie and overlap of the neighbors of nodes at both the ends indicate an existence of weak ties in the network. 
If a link has a high betweenness centrality $(\beta_{L})$,
it is known to be stronger as it helps in connecting different modules in the network.
These weak ties are cited to be important in connecting different communities \cite{edbc_ol}. 
For PPI networks considered here, we find negative $O-\beta_{L}$ correlation for all 
the normal and disease networks 
which suggests the presence of weak ties in these networks.
This is in line of the earlier observation that the PPI networks are known to compose of 
different metabolic cycles and biological pathways. A protein involved 
in particular pathway plays role in regulating other pathways as well, termed as cross talk between pathways \cite{SubGraphs, SubGraphs1}.

\begin{figure}[ht]
\begin{center}
\includegraphics[width=12cm,height=6cm]{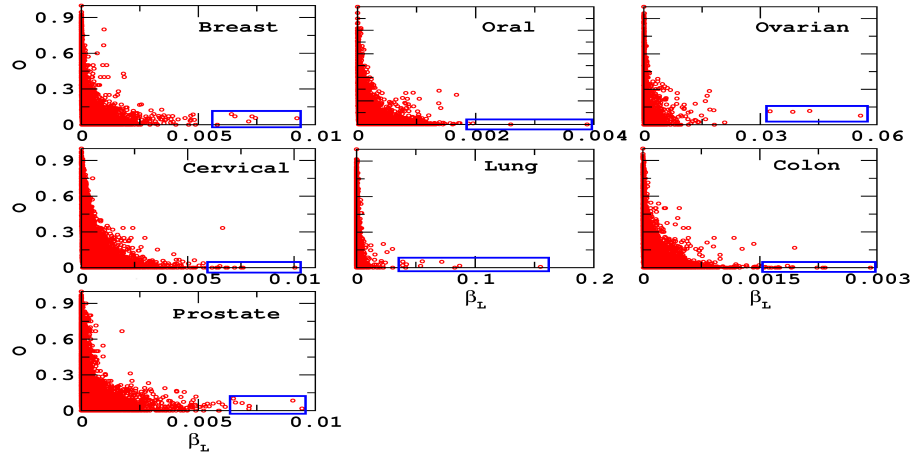}
\end{center}
\caption{{\bf Weak ties analysis.} The $O-\beta_{L}$ analysis for all the disease networks reveal negative correlation (red circles). We highlight the edges (blue box) having high $\beta_{L}$ and low $O$ in all the disease and find the presence of 63 common proteins in those edges.}
\label{fig_bc-ol}
\end{figure}

The weak ties analysis reveals 122 proteins (61 pairs) for all the disease networks together.
(Fig. \ref{fig_bc-ol} blue box). Among the 63 disease common proteins, 
ten proteins possess the properties of weak ties. Of these ten proteins, five proteins namely MUC1, SOD2, MAPK1, HSPA4 and HSPA5 
are among the six proteins which we have already listed for their importance 
in possessing the property of high $\beta_c$ and low $k$. This is not surprising
as the nodes having high betweenness and low $k$ are highly probable of having less overlap
and thus the link betweenness of those nodes become high.
The proteins having both the weak ties as well as high $\beta_{c}$ low $k$ properties (Fig. \ref{Fig_Bits_1})
indicates participation of these proteins in many pathways in the network
as well as their significant role in causing cancer which we discuss in the following section.

We discern the functional characterization 
of the proteins revealed in the above analysis
based on their sub-cellular locations i.e. sensors and effectors. 
The sensors and effectors are widely characterized in 
direct or indirect involvement of a protein in cancer biology.
Further, it is reported that post transcription regulators such as non-coding RNAs, 
particularly the miRNAs effectively regulate the expression of sensors \cite{mirna0}.
miRNAs are a class of short non-coding RNAs with post transcription regulatory functions.  
Here, we study the role of miRNAs to understand the regulation of these proteins at the transcription level.

\subsection*{Functional role and miRNA analysis of important proteins}

\begin{figure}[ht]
\begin{center}
\includegraphics[width=8cm,height=4cm]{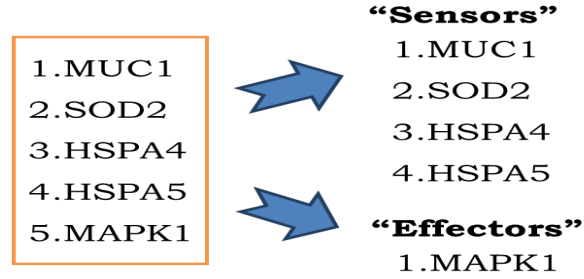}
\end{center}
\caption{{\bf  
Proteins with high $\beta_{c}$-low$k$ and weak ties.} MUC1, SOD2, HSPA4 and HSPA5 are sensors having role as upstream components of intra-cellular signaling cascades whereas MAPK1 is a downstream molecules classified here as effector. }
\label{Fig_Bits_1}
\end{figure}

We find that out of five proteins participating in the weak ties as well as having high $\beta_{c}$ low $k$ properties, 
four proteins MUC1, SOD2, HSPA4 and HSPA5 are under the category of sensors and 
one protein MAPK1 is categorized under effectors (Fig. \ref{Fig_Bits_1}). 
The proteins marked under `sensors' category are primarily upstream components of intra-cellular signaling cascades, 
altered expression of which may lead to downstream activities in the tumor milieu \cite{Tabassum}. 
The protein under `effector' category is downstream molecule which is often implicated but is not exclusive to cancer and therefore, for  elaborative studies, we only discuss sensors (Fig. \ref{Fig_Bits_1}). 

\begin{figure}[ht]
\begin{center}
\includegraphics[width=11cm,height=11cm]{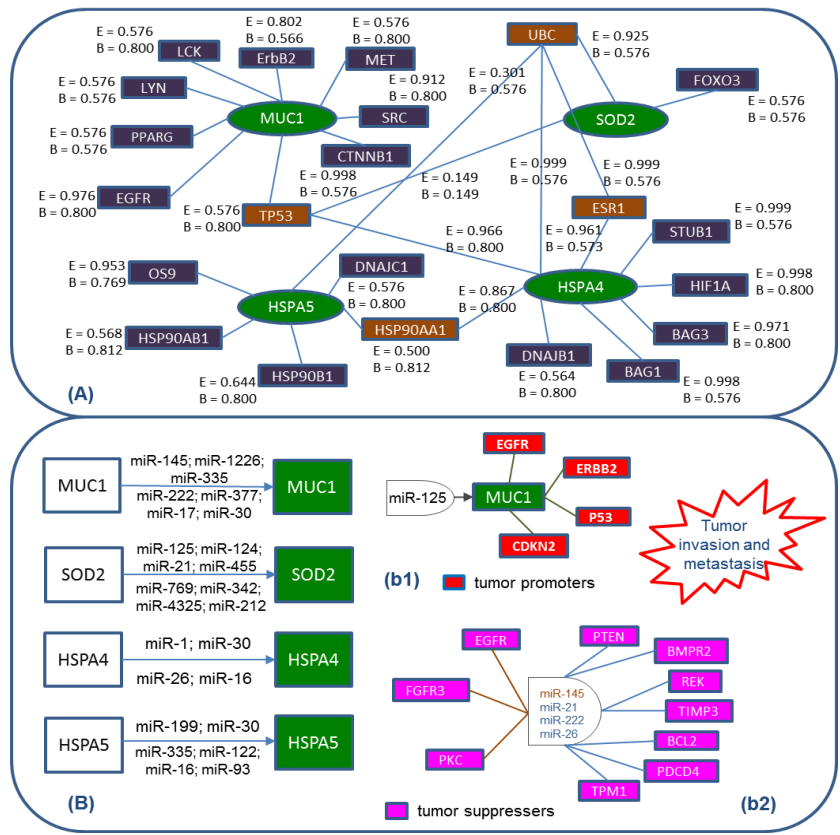}
\end{center}
\caption{{\bf Sensor proteins and their miRNAs associations.} (A) depicts the interactions (blue boxes) of the 
sensor proteins (green elipse) from STRING database. 
Red boxes are the common neighbors of the four sensor proteins.
(B) depicts the miRNAs associated with the sensor proteins. miRNA in (b1) is a positive regulator of MUC-1 leading to 
down regulation of tumor promotors (red boxes) while, other miRNAs in (b2) participate in up-regulation of MUC1, SOD2 and HSPA4 
and down-regulate the activities of tumor suppressors (pink boxes). }
\label{fig_mirna}
\end{figure}

\paragraph*{Functional role:} 
We first scan the significant interactions of these proteins from the STRING database,
which is based on probabilistic confidence score ($>0.50$). 
The associations in STRING are based on high throughput experimental data, 
thorough search of the databases and predictions based on genomic context analysis. 
Thereafter, utilizing and incorporating information of interacting partners by KEGG pathway analysis, 
we find that all these sensor proteins HSPA4, HSPA5, MUC1 and SOD2 have very important interacting proteins 
e.g. ESR1, ErBb2, UBC, OS9 etc that have significant and specific 
involvement in the pathways participating in proliferation and migration of cancerous cells (Fig. \ref{fig_mirna}(A)).
Moreover, even after removing lower probabilistic interactions, 
these proteins possess some common neighbors viz. TP53, HSP90AA1, ESR1 and UBC which 
illustrates that these proteins are interlinked to each other and 
suppressing the expression of any of these may lead to the blockage of 
the cancer associated pathways (Fig. \ref{fig_mirna}(A) red boxes). 
Further, we look into the miRNA molecules associated with these proteins 
that can help regulate the expression of these sensor proteins.

\paragraph*{miRNA analysis:}

Recent studies have shown that the expression of miRNAs is de-regulated in cancer progression, 
tumor invasion, metastasis, and subsequent chemoresistance \cite{mirna1,mirna2}. 
For the four sensor proteins, we extract the list of experimentally validated miRNAs 
from Sanger Institute’s miRTarBase database regulating the sensors proteins \cite{tarbase}.
These miRNAs are then filtered for their role in regulating both the 
sensor proteins as well as their interacting partners (Fig. \ref{fig_mirna} (B)). 
Here, we present example of few miRNAs corresponding to each sensor protein to demonstrate how miRNAs can be used to find out essential 
protein biomarkers and their downstream pathway roles.
For instance, based on our studies, we find miR-125b as a probable miRNA 
regulating both MUC-1 and its interacting partners like, EGFR, ERBB2, CDKN2A which are potent tumor promoters 
(Fig. \ref{fig_mirna} (B-b1)). 
Interestingly, KEGG pathway analysis, depicting various miRNAs de-regulated in cancer, reports that miR125 is down-regulated in various cancers which provides strong indication 
to the consequences of over-expression of MUC-1 in cancer (Fig. S6). 
Similarly, miR145 is predicted to have potential binding affinity for MUC1 which 
is suppressed in many cancers and is reported to have targets 
like, EGFR, FGFR3 and PKC (Fig. \ref{fig_mirna} (B-b2)) which have well established role in tumorigenesis \cite{mirna_rajdeep}. 
Further, miRNAs like, miR-21, miR-26 and miR-222 are up-regulated in various cancers 
and found to regulate MUC1, SOD2 and HSPA4. 
Existing literature suggests that these miRNAs also down-regulate tumor suppressors 
like, PTEN, BMPRII, RECK, TIMP3, BCL2, PDCD4, TPMI thereby promoting cancer (Fig. \ref{fig_mirna} (B-b2)). 
Further, to have a complete idea about miRNA-mediated regulation, 
we calculate the probabilistic distribution of proteins regulated by a given miRNA 
which also controls the expression of sensor proteins and study the role these miRNAs play in regulating other proteins.
It is revealed that the MAPKinase family is highly probable target (Fig. S9) 
since they are simultaneously being regulated by majority of miRNA which regulate sensor proteins. 
The implication of this investigation is that the proteins of the particular signaling pathway is highly important in cancer dataset under study 
and it can be chosen as a suitable target to be looked upon after miRNA inhibition. 
Apart from the MAPKs, other common targets are HSPs, B-catenin, PI3K/Akt, Mucin family,
which is based on the probability scores alone (Table. S10). 
In all, the data indicates the merits of using network theory to predict plausible proteins regulating a range of downstream targets. However, experimental validation is essential for a concrete conclusion.

\section*{Conclusion}
We analyze the protein-protein interaction networks of the 
normal and the disease conditions of seven different cancers under the 
combined framework of the spectral graph theory, 
network theory along with the multilayer analysis. 
The analysis exhibits overall similar behavior of various structural properties 
among the normal and the disease states (with few exceptions) such as a high clustering coefficient,  
small diameter, negative $k-CC$ correlation and positive degree-degree correlation.
Further, these properties exhibit significant deviations from those of
 their corresponding random networks.
While, high $\langle CC \rangle$ and negative $k-CC$ correlation depicts 
complexity in the underlying system, change in the values of $r$ and $\langle CC \rangle$
from the normal to the disease states signify changes in the interaction patterns between the datasets.
Further, the $r$ and $\langle CC \rangle$ values exhibit unsystematic changes 
from the normal to the disease networks, 
i.e. while some of the cancer networks have higher values of $r$ and $\langle CC \rangle$ than their normal counterpart, 
some have lower values as compared to the normal depicting inherent complexity in the normal and the disease networks.

Furthermore, to have a deeper insight into the complexity of the normal and the disease system
and have in-depth analysis of the factors leading to changes in the interaction pattern in the networks, 
we analyze spectra of the cancer and normal networks, as well as compare them with those of their respective random models. 
We find that there is a high degeneracy at the zero and the minus one eigenvalue.
The zero degeneracy is directly related with the number of duplicate nodes in the network 
and occurrence of different duplicate nodes in the normal and disease states suggest evolutionary 
changes from the normal to the disease state. Whereas, a
high degeneracy at minus one eigenvalue suggests abundance of complete sub-graphs in the 
networks which may be important for proper functioning of the underlying system. 
Moreover, difference in the behavior of prostate cancer for various structural and spectral properties 
than those of the other cancer datasets may be due to incomplete knowledgebase of proteomic 
interactions or may have due to independent cancer development processes for this cancer.

Next, the multilayer framework reveals 63 common proteins among all the disease datasets. 
These 63 proteins show much higher $\langle CC \rangle$ as compared to that of the whole networks.
Further, the functional analysis of these 63 proteins through pathway ontology reflect 
their involvement in 
important cancer related pathways such as Vascular endothelial growth factor (VEGF) signaling, 
PIP3 activation, VEGFA-VEGFR2 pathway, signaling by PDGF, signaling by NGF etc.

Other network properties investigated for the 63 proteins common to all the 
diseases namely the $k-\beta_{c}$ correlation 
and weak ties ($O-\beta_{L}$) analysis highlights few proteins and their links 
that are structurally important in the network. These proteins are also found to have
functional significance responsible for the occurrence of cancer.
These proteins are further categorized into sensors and effectors. 
The sensors having primary role in contributing changes in the tumor \cite{Tabassum}, 
are studied for post transcription regulators such as the miRNAs 
as they effectively regulate the expression of sensor proteins.
Out of 5 common proteins which posses structural importance in the individual
networks, 4 are sensor
proteins.
The miRNA study of these sensor proteins reveals their involvement in
tumor invasion and metastasis thereby suggesting their role in progression of the cancer. 
Further experimental validations of these miRNAs can help in making corresponding
 proteins as potential drug targets.

All these results based on rigorous analysis using sophisticated mathematical 
and statistical technique along with the extensive data collection and functional literature 
survey enables us to understand various cancers at the fundamental level.
The framework considered here focuses on finding important proteins based on their position in the individual networks, which 
can be extended to those diseases for which very less information is available about the genes which are responsible for the occurrence of the disease. Furthermore, 
multilayer framework revealing common proteins for different cancers 
provide a direction for developing novel drugs, therapeutic targets and biomarkers along with the nascent concept of single drug therapy for multiple diseases and personalized medicine in a time efficient and cost effective manner.

\section*{Methods}
\subsection*{Data assimilation and Network construction}
There are two basic components of a network namely, nodes and edges. 
Here we study PPI networks of the normal and the disease cells 
where nodes are the proteins and edges denote interactions between the proteins. 
Nodes in a normal and the corresponding disease network are selected on the basis of their expression 
in a cell of the normal or disease tissue, respectively.
For instance if a protein is expressed as in the normal state of the breast cell, it is considered in the construction of the breast normal network and similarly, if a protein is expressed in the breast tumor (malignant) cell, it is considered in construction of the breast cancer network.
After diligent and enormous efforts of mining literature and database text, we collect the list of proteins in the normal tissues and the corresponding cancer tissues from various literature and bioinformatics sources (databanks) namely GenBank \cite{Genbank} and UniProtKB, which mines
the proteomic data from various other repositories like European Bioinformatics Institute, the Swiss Institute of Bioinformatics, and the Protein Information Resource etc \cite{uniprot}. 
We enlist the proteins for a particular tissue by searching relevant keywords, such as its target tissue/cell type in the search panel of various databanks.
To keep the authenticity of the data we only take those proteins into account which are reviewed and cited (literature authenticated). 
Additionally, there are numerous cell lines available for biological studies, but a very few have been exploited for their maximum proteomic insight. We gather the protein expression informations through various
cell-lines comprising of different origin (human, mouse, horse etc) from available online literatures to make the data more complete. The details of the cell-lines databanks can be found in the ``Data collection and network construction" sub-section of the supplementary material.
The details of all the proteins for seven different tissues for the normal
and disease states can be accessed from \cite{data}.
Once all the proteins for seven different tissues for the normal and disease states are collected, leading to fourteen datasets, the interacting partners of these proteins are retrieved from the STRING database version 9.1 \cite{STRING}. A interaction between a pair of proteins is considered if there exists a direct (i.e. physical), indirect (i.e. functional) or both relation between them. STRING provides the physical and functional interactions for a given list of the proteins. Note that in this work, we consider interactions between a pair of proteins from
the STRING database for both the normal and disease networks so any change in the interactions in a particular state (disease or normal) arises only due to deletion or addition of new nodes in the networks. Note that, the information on whether a pair of proteins are really interacting in a particular state, i.e. the dynamical nature of PPIs are missing.
In this way, we have seven networks for the normal and seven networks for the corresponding disease states. The protein-protein interactions of all the fourteen networks as adjacency list can be found in \cite{data}.

Next, we define the interaction matrix or the adjacency matrix ($A_{\mathrm {ij}}$) of the network as,
\begin{equation}
A_{\mathrm {ij}} = \begin{cases} 1 ~~\mbox{if } i \sim j \\
0 ~~ \mbox{otherwise} \end{cases}
\label{adj_wei}
\end{equation} 
We investigate the PPI networks for their various structural and spectral properties.

\subsection*{Properties of Complexome}
\paragraph*{Structural properties}
Several statistical measures have been proposed to understand specific features of a network \cite{Barabasi_2002, bocaletti}. 
One of the most basic structural parameter of a network is the degree of a node ($k_i$),
which is defined as the number of neighbors of a node has ($k_i=\sum_j A_{ij}$). 
The degree distribution $P_{k}$, revealing the fraction of vertices's with the degree $k$, 
is known to be a fingerprint of the underlying network \cite{Barabasi_2002}. 
Another important parameter of a network is the clustering coefficient $(CC)$ of the network. 
Clustering coefficient ${CC}_i$ of a node (say $i$) can be written as the ratio of the number of interactions the neighbors of a particular node has and number of possible connections the neighbors can have \cite{Watts, newman}.
The average of all the individual $CC_i$ gives the average clustering coefficient ($\langle CC \rangle$). 
It characterizes the overall tendency of the nodes to form clusters or groups.
Further, the betweenness centrality ($\beta_c$) 
of a node $i$ is defined as the fraction of shortest paths that pass through the node $i$ \cite {Barabasi_2002}, 
\begin{equation}
\beta_{c}^{i}=\sum_{st} \frac{n^i_{st}}{g_{st}},
\end{equation} 
$n^i_{st}$ denotes the number of paths from $s$ to $t$ that passes through $i$ and $g_{st}$ 
is the total number of paths from $s$ to $t$ in the network. 
Further, we calculate the diameter of the network which measures the longest of the shortest paths 
between all the pairs of the nodes \cite{diameter}.

Another important property of a network that helps us in distinguishing the normal from the disease datasets is degree-degree correlation ($r$).
This property measures the tendency of nodes to connect with the nodes having similar number of edges \cite{Alok_r, newman_r, Rivera} and can be defined as, 
\begin{equation}
r = \frac{[\frac{1}{M} \sum_{i} j_i k_i] - [ \frac{1}{M} \sum_i \frac{1}{2}(j_i + k_i)^2]}
{[ \frac{1}{M} \sum_{i} (j_i^2+ k_i^2)] - [ \frac{1}{M} \sum_i \frac{1}{2}(j_i + k_i)^2]}, 
\label{assortativity}
\end{equation}
where $j_{i}$ and $k_{i}$ are the degrees of the nodes connected through the $i^{th}$ edge, and $M$ is the total number of edges in the network.
The value of $r$ being negative (positive) corresponds to a dis(assortative) network.

Further, to understand the network architecture, we identify the weak ties in the network 
by calculating the edge-betweenness centrality and overlap of the pair of nodes. 
Granovetter's weak ties hypothesis: a socially driven network tool 
highlights the importance of an edge in the network through the strength of their tie by
calculating the edge-betweenness centrality ($\beta_{L}$) with inverse relation
to the overlap ($O$) of their neighborhoods \cite{Granovetter}.
The $\beta_{L}$ can be defined as, 
$ \beta_{L} = \sum_{v \in V_{s}} \sum_{w \in V/{v}} \sigma_{vw} (e)/\sigma_{vw}$,
 where $\sigma_{vw} (e)$ is the number of shortest paths between $v$ and $w$ that contain $e$, and $\sigma_{vw}$ is the total number of shortest paths between $v$ and $w$ \cite{edbc_ol}. 
Next, the overlap of the neighborhood ($O_{ij}$) of two connected nodes $i$ and $j$ is defined as,
$O_{ij} = \frac{n_{ij}} {(k_{i} - 1) + (k_{j} - 1) - n_{ij}}$, where $n_{ij}$ is the number of neighbors common to both nodes $i$ and $j$ \cite{edbc_ol}. 
Here, $k_i$ and $k_j$ represent the degree of the node $i$ and $j$.
Then, we calculate Pearson correlation coefficient ($O-\beta_{L}$) of $O_{ij}$ and $\beta_{L}$ as,
\begin{equation}
O-\beta_{L}= \frac{( O_{ij}-\langle O_{ij}\rangle ) (\beta_{L} - \langle\beta_{L}\rangle )}{\sqrt{( O_{ij}- \langle O_{ij}\rangle )^2} \sqrt{(\beta_{L} - \langle \beta_{L} \rangle)^2}}
\label{lap_eigval_entp}
\end{equation}

\paragraph*{Spectral properties}
Let us denote eigenvalues of the adjacency matrix by $\lambda_i$, $i = 1, 2, \ldots , N$ such that $\lambda_1$ $ < $ $\lambda_2$ $ < $ $\lambda_3$ $ <  \ldots  <$ $\lambda_N$.
Further, in order to understand the evolutionary mechanisms involved in normal and cancer state, that plays an important role in the formation of these PPI networks, we calculate the degenerate eigenvalues in the network.
First, we investigate the role of node duplication by identifying the nodes sharing exactly the same neighbors from the corresponding adjacency matrices \cite{Alok, Kitano}.
When (i) two rows (columns) have exactly same entries, it is termed as complete row (column) duplication $R_1 = R_2$, (ii) the partial duplication of rows (columns) where $R_{1} = R_{2} + R_{3}$, where, $R_{i}$ denotes $ith$ row of the adjacency matrix. 
The count of zero eigenvalues ($\lambda_0$) provides an exact measure of (i) and (ii) conditions \cite{Golub}.
Further, we calculate degeneracy at minus one eigenvalues ($\lambda_{-1}$) which 
provides an insight to the complete sub-graphs in the network \cite{spectra_book}.

\subsection*{Multilayer Framework}

Analysis of structural and spectral properties suggests overall similarities between the normal and the disease PPI networks.
All the seven disease networks are represented as different
layers of a disease multilayer network. Similarly, all the seven normal networks form different
layers of a normal multilayer network leading to the normal multilayer network framework. We extract common nodes from (i) all the normal networks, (ii) all the disease networks and (iii) common between all the disease and all the normal networks (union of (i) and (ii)), and investigate various structural and functional properties of the common proteins referring it as multilayer analysis of these three independent subtractive PPI networks for each cancer (Fig. \ref{Fig2_multilayer}(a)).
After extracting common proteins, we find their interacting partners from all the disease datasets and analyze various properties of those proteins which are common in all the disease networks.

\subsection*{Construction of Erd\H os R\' enyi and Configuration networks }

Further, we compare properties of PPI networks with the corresponding ER random networks having the same average degree $\langle k \rangle$ and size $N$ \cite{Barabasi_2002}, where nodes are randomly
connected with a fixed probability p, calculated as k/N. This leads to ER networks having the same density distribution as of the corresponding real networks. We use ER network ensemble of 20 networks for all the properties discussed here.

Additionally, we compare properties of PPI normal and disease networks with the corresponding configuration networks. The configuration model in addition of having the same size and number of connections as of a given network, preserves the degree sequence of the given network,
by generating a random network with a given degree sequence of an array of size $m=\frac{1}{2}{\sum_{i=1}^Nk_i}$. 
We construct the configuration model network by
taking the degree sequence of various PPI networks as
input. Each node of the corresponding configuration model
is allotted stubs equal to their degree, and then these stubs
are paired with a uniform probability \cite{config, config1}. This generates a configuration model for a given degree sequence.
We generate 20 such realizations for a given degree sequence.

\section*{Acknowledgments}
SJ acknowledges Council of Scientific and Industrial Research (CSIR) grant (25(0205)/12/EMR-II) 
and Department of Science and Technology (DST), Government of India
grant EMR/2014/000368 for financial support.
AR thanks Amit Kumar Pawar and Sanjiv Kumar Dwivedi for helping with data and codes, respectively and 
all the members of Complex Systems Lab for useful discussions.

\section*{Author contributions}
SJ conceived and supervised the project. AR collected the data and constructed the networks. 
SJ, AR and PP performed the numerical experiments and analyzed the data for various network properties. 
AR, JN and KL did the functional analysis of the important proteins. JN, KL and RC performed the miRNA analysis.
All the authors wrote and approved the manuscript.

\section*{Additional information}
\subsection*{Supplementary Information} accompanies this paper at http://www.nature.com/naturescientificreports
\subsection*{Competing financial interests:} The authors declare no competing financial interests.


\newpage

\section*{Figure legends}

\newpage

\newpage

\section*{Tables}[ht]
\begin{table}[ht]   
\begin{tabular} {p{0.3cm} p{5.0cm} p{10cm}}  \hline
{\bf Sr.}& {\bf Pathway} & {\bf Proteins involved}\\\hline

1. & Signaling by Vascular endothelial growth factor (VEGF) & IQGAP1, FN1, PTK2, AKT1, FGFR2, MAPK1, VEGFA, CTNNA1, CTNNB1, CAV1\\ 
2. & Signaling by Stem cell factor receptor (SCF-KIT) &	IQGAP1, FN1, STAT1, GSK3B, PTK2, AKT1, FGFR2, MAPK1, PTEN	\\ 
3. & VEGFA-VEGFR2 (VEGF family receptors) Pathway	&	IQGAP1, FN1, PTK2, AKT1, FGFR2, MAPK1, VEGFA, CTNNA1, CTNNB1, CAV1	\\ 
4. & Signaling by epidermal growth factor receptor 4 (ERBB4) &	IQGAP1, FN1, ESR1, GSK3B, PTK2, AKT1, FGFR2, MAPK1, PTEN	\\ 
5. & Protein kinase B (AKT) signaling  &GSK3B, AKT1, FGFR2, PTEN	\\
6. & Cellular responses to stress	&	HSPA4, PRDX5, EP300, GSK3B, SOD2, MAPK1, NBN, VEGFA, HSPA5, PRDX2, PRDX1	\\ 
7. & Signaling by Platelet-derived growth factor (PDGF)	&	IQGAP1, FN1, STAT1, GSK3B, PTK2, AKT1, FGFR2, MAPK1, PTEN	\\
8. & Downstream signaling of activated FGFR2	&	IQGAP1, FN1, GSK3B, PTK2, AKT1, MAPK1, FGFR2, PTEN	\\ 
9. & Signaling by Nerve Growth Factor (NGF) &	IQGAP1, FN1, GSK3B, PTK2, AKT1, ARHGDIA, MAPK1, FGFR2, PTEN, RTN4	\\ 
10. & Signaling by Rho GTPases	&	IQGAP1, BIRC5, CDH1, PTK2, ARHGDIA, MAPK1, SFN, CTNNA1, CTNNB1, CTTN	\\
11. & Axon guidance	&	IQGAP1, FN1, GSK3B, PTK2, MMP2, FGFR2, MAPK1, VEGFA, CFL1	\\ 
12. & Innate Immune System	&	IQGAP1, FN1, EP300, GSK3B, PTK2, AKT1, FGFR2, MAPK1, PYCARD, PTEN, CFL1, CTNNB1	\\ 
13. & Signal Transduction	&	IQGAP1, FN1, STAT1, EP300, GSK3B, ARHGDIA, SFN, CTNNA1, CTNNB1, CAV1, BIRC5, ESR1, CDH1, PTK2, AKT1, MAPK1, FGFR2, VEGFA, PTEN, RTN4, CTTN	\\ 
14. & Metabolism of proteins	&	PDIA3, LMNA, PABPC1, BRCA1, MMP2, HSPA5, MUC1, CTNNB1, RPS3, PML	\\ \hline
\end{tabular}
\caption{{\bf Molecular and pathway ontology of 63 common proteins.} Set of proteins are involved in particular cellular pathway having
major role in occurrence of different types of cancers.}
\label{63_prtns}
\end{table}

\end{document}